\documentclass[12pt,epsf]{article}
\usepackage{amsmath}
\usepackage{amssymb}
\usepackage{bm}
\usepackage[dvips]{graphics}

\newcommand{\be}{\begin{equation}}
\newcommand{\ee}{\end{equation}} 
\newcommand{\bea}{\begin{eqnarray}}
\newcommand{\eea}{\end{eqnarray}}

\newcommand{\cV}{{\cal V}}

\newcommand{\ad}{\mbox{ad}\,}

\def\tr{\mbox{tr}}
\def\Tr{\mbox{Tr}}
\def\Str{\mbox{Str}}

\begin{document}
\setlength{\oddsidemargin}{0cm}
\setlength{\baselineskip}{7mm}

\begin{titlepage}  \renewcommand{\thefootnote}{\fnsymbol{footnote}}
$\mbox{ }$
%\vspace{-3cm}
\begin{flushright}
\begin{tabular}{l}
KUNS-1884 \\
hep-th/0312026\\
December 2003
\end{tabular}
\end{flushright}

~~\\
~~\\
~~\\

\vspace*{0cm}
    \begin{Large}
       \vspace{2cm}
       \begin{center}
         {Supersymmetric large-$N$ reduced model with multiple matter}
\\
       \end{center}
    \end{Large}

  \vspace{1cm}

\begin{center}

          Hikaru K{\sc awai}$^{ab}$\footnote
           {
E-mail address : hkawai@gauge.scphys.kyoto-u.ac.jp},
          Tsunehide K{\sc uroki}$^a$\footnote
           {
E-mail address : kuroki@gauge.scphys.kyoto-u.ac.jp}{\sc and}
          Takeshi M{\sc orita}$^a$\footnote
           {
E-mail address : takeshi@gauge.scphys.kyoto-u.ac.jp}

$^a$           {\it Department of Physics, Kyoto University,
Kyoto 606-8502, Japan}\\
$^b$           {\it Theoretical Physics Laboratory, RIKEN (The
  Institute of Physical and Chemical Research), Wako, Saitama 
351-0198, Japan}\\
\end{center}

\vfill

\begin{abstract}
\noindent 
We showed in hep-th/0303210 that the Dijkgraaf-Vafa theory can be regarded as large-$N$ reduction 
in the case of  $\mathcal{N}=1$ supersymmetric $U(N)$ gauge theories, with single adjoint matter.
We generalize this to gauge theories with gauge groups being the products of some unitary  groups coupled to bifundamental or fundamental matter.
We show that some large-$N$ reduced models of these theories are supermatrix models, whose free energy is equivalent to the prepotentials of the original gauge theories.
The supermatrix model in our approach should be taken in the Veneziano limit $N_c,N_f \rightarrow \infty $ with $N_f/N_c$  fixed.

\end{abstract}

\vfill
\end{titlepage}
\vfil\eject

\section{Introduction}
In the theory of Dijkgraaf-Vafa \cite{DV} \cite{CDSW}, $\mathcal{N}=1$ supersymmetric gauge theory is  related to a large-$N$ bosonic matrix model .
They have discovered that the prepotential of the gauge theory is the free energy of the matrix model.

We have shown that this result can be regarded as a large-$N$ reduction \cite{EK} \cite{KKM}.
One can map the gauge theory on non-commutative space to a large-$N$ reduced model (see the Appendix) 
and  show that the holomorphic parts of the reduced model (supermatrix model) reproduce the prepotential of the gauge theory as the free energy.

However, we have only demonstrated the Dijkgraaf-Vafa theory via the reduced model in the simplest system.
There are many applications of the Dijkgraaf-Vafa theory to other gauge groups and the inclusion of matter \cite{W} \cite{S} \cite{Hof},
and we want to understand such applications in terms of reduced models.
Such understanding is not only important in respect of the Dijkgraaf-Vafa theory itself, but also gives  insight for constructing a general connection between field theory and matrix models.
For example, we have not yet found a reduced model for $SO(N)$ Yang-Mills theory, while  
 the matrix model for this gauge theory is given in the Dijkgraaf-Vafa theory \cite{W}.
This will help us to construct its reduced model.
Thus, understanding these theories is important in the context of large-$N$ reduction.

We generalize in this paper the argument given in \cite{KKM} to gauge theories which have gauge groups of the products of some unitary groups and are coupled to their adjoint, bifundamental and fundamental matter.
The low-energy effective action of these theories is given by the  prepotential,
and this prepotential is equivalent to the free energy of the corresponding matrix models.

We first apply our approach to  a gauge theory which has $n$ gauge groups,
\begin{align*}
\prod_{i=1}^{n}U(N_i),
\end{align*}
 coupled to $n$ adjoint, bifundamental and  anti-bifundamental matter.
We show that the reduced model of this theory reproduces the matrix model in the Dijkgraaf-Vafa theory.

We next consider the $A_2$ quiver-like theory which has two gauge groups
\begin{align}
U(n_c)\times U(n_f).
\label{A_2}
\end{align}
We regard the $U(n_f)$ gauge group as a global symmetry by setting vector superfield $V_f$ for the $U(n_f)$ gauge group to $0$.
This system then becomes $U(n_c)$ gauge theory with $n_f$  fundamental and anti-fundamental matter \cite{Hof}.
This theory is  the first example in which the large-$N$ reduction reproduces the Dijkgraaf-Vafa theory by taking $V_f=0$
.

We finally consider a general system which has several unitary gauge groups, flavor symmetries and  matter in the adjoint, 
bifundamental and fundamental representations.
We show that the Dijkgraaf-Vafa theory of this system can be reproduced by its reduced model, as in the previous two cases.\\

In section \ref{2} of this paper, we discuss the application of the gauge theory with bifundamental matter in our approach.
In section \ref{3}, we consider a system with fundamental matter and a more general system. 
Section \ref{4} contains conclusion and a discussion.
The Appendix presents a brief review of non-commutative space and  the construction of the supersymmetric reduced model of $U(n)$ gauge theory coupled to adjoint matter.

\section{Supersymmetric reduced model with
 bifundamental matter}
\label{2}
We demonstrate in this section the Dijkgraaf-Vafa conjecture of  the gauge theory with bifundamental matter in terms of large-$N$ reduction.
As we mentioned in the introduction, the argument is relevant to the single adjoint matter case \cite{KKM}.
We first consider this theory on bosonic non-commutative space and construct a supersymmetric reduced model.
We next introduce fermionic non-commutative space and map the supersymmetric reduced model to a supermatrix model.
We anticipate that the holomorphic parts of this supermatrix model will reproduce those of the gauge theory when we set the non-commutativity 
$C^{\mu\nu},\gamma^{\alpha\beta}$ to $0$.
We indeed check that the equivalence between the free energy and prepotential 
holds in this theory.

\subsection{Construction of the supersymmetric reduced model}
\label{2.1}
\setcounter{equation}{0}
We consider the gauge theory with multiple bifundamental matter. 
This theory comprises gauge groups
\begin{align}
\prod_{i=1}^{n}U(N_i),
\end{align}
and matter.
We denote  chiral matter in the adjoint representation of $U(N_i)$ as $\Phi_{i}$, a vector superfield as $V_{i}$, and chiral matter in the bifundamental representation of $U(N_i)$ and $U(\bar{N}_j)$ $(i>j)$ as $X_{i,j}$ and anti-bifundamental representation as $Y_{j,i}$,
where index $i$ of the bifundamental representation runs from 1 to $n$. 
Field strength $W_{i \alpha}$ for each $U(N_i)$ is defined as
\begin{align}
W_{i \alpha}=-\frac{1}{4}\bar{D}\bar{D}e^{-V_{i}}D_\alpha e^{V_{i}}.
\end{align}
With these fields, we consider the following action:
\begin{align}
S= & \int d^4 x d^2 \theta d^2 \bar{\theta} \Biggl\{ ~\sum_{i=1}^n\tr_{U(N_i)} ( \bar{\Phi}_i e^{V_i} \Phi_{i} e^{-V_i}) \nonumber \\
&+\sum_{i>j} \tr_{U(N_j)}(\bar{X}_{i,j} e^{V_i} X_{i,j} e^{-V_j}) 
+\sum_{i>j} \tr_{U(N_j)}(\bar{Y}_{j,i} e^{V_j} Y_{j,i} e^{-V_i})\Biggr\} \nonumber \\
&+ \int d^4x d^2 \theta\Biggl\{~\sum_{i=1}^n 2\pi i \tau_i \tr_{U(N_i)}W_i^\alpha W_{i\alpha}  
+ \sum_{(i,a)} g_{(i,a)} \tr_{U(N_i)} O_{(i,a)}(\Phi,X,Y)\Biggr\} +c.c.,
\label{action1}
\end{align}
where $\tau_i$ is the gauge coupling constant.

The superpotential, 
\begin{align*}
\sum_{(i,a)} g_{(i,a)} \tr_{U(N_i)} O_{(i,a)}(\Phi,X,Y)
\end{align*}
is a single-trace polynomial in which subscript $i$ indicates a gauge group and subscript $a$ indicates a form of the function $O_{(i,a)}$.
We can explicitly express the polynomial as
\begin{align}
\sum_{(i,a)} g_{(i,a)} \tr_{U(N_i)} O_{(i,a)}(\Phi,X,Y)
=& \sum_{i=1}^n \sum_{k=1}^{n_i}\frac{1}{k+1}a_{(i,k)} \tr_{U(N_i)} \Phi_i^{k+1}\nonumber \\
&+ \sum_{i,j,k}b_{(i,j,k)}~ \tr_{U(N_i)} Y_{i,j} \Phi^k_j X_{j,i} \nonumber \\
&+ \sum_{i,j,p,q}c_{(i,j,p,q)}~ \tr_{U(N_i)} Y_{i,j} \Phi^p_j X_{j,i} Y_{i,j} \Phi^q_j X_{j,i} + \cdots,
\end{align}
where $a_{(i,k)}, b_{(i,j,k)}, c_{(i,j,p,q)}, \cdots$ are coupling constants.
We impose the one condition on the superpotential that all the adjoint matter is massive, 
because massless adjoint matter gives singularity in low-energy theory and may  break the holomorphy.\\

We consider this theory on bosonic non-commutative space (\ref{bosc}) and construct a reduced model of the theory \cite{NCYM}.
As in \cite{KKM}, a reduced model of this gauge theory is given by
\begin{align}
S=& 
 (2\pi)^2 \sqrt{\det C}\int  d^2 \theta d^2 \bar{\theta}~\Biggl\{ \sum_{i=1}^n \Tr_{U(\hat{N}_i)} ( \hat{\bar{\Phi}}_i e^{\hat{V}_i} \hat{\Phi}_{i} e^{-\hat{V}_i}) \nonumber \\
&+\sum_{i>j} \Tr_{U(\hat{N}_j)}(\hat{\bar{X}}_{i,j} e^{\hat{V}_i} \hat{X}_{i,j} e^{-\hat{V}_j}) 
+\sum_{i>j} \Tr_{U(\hat{N}_j)}(\hat{\bar{Y}}_{j,i} e^{\hat{V}_j} \hat{Y}_{j,i} e^{-\hat{V}_i}) \Biggr\} \nonumber \\
&+ (2\pi)^2 \sqrt{\det C} \int d^2 \theta~ \Biggl\{ \sum_{i=1}^n 2\pi i \tau_i \Tr_{U(\hat{N}_i)}\hat{W}_i^\alpha \hat{W}_{i\alpha} \nonumber \\
&+ \sum_{(i,a)}g_{(i,a)}\Tr_{U(\hat{N}_i)}O_{(i,a)}\left( \hat{\Phi}(\theta), \hat{X}(\theta),\hat{Y}(\theta) \right)  \Biggr\}+ c.c..
\label{reduced}
\end{align}
Here $\hat{\Phi}_i(\theta)$, $\hat{X}_j(\theta)$, $\hat{Y}_k(\theta)$ and $\hat{V}_l(\theta,\bar{\theta})$ are infinite matrices whose components are a function of $\theta$ and/or $\bar{\theta}$, $\hat{N}_i$ is their infinite rank, and  
\begin{align}  
\hat{W}_{i\alpha} (\theta)
&=-\frac{1}{4}\frac{\partial}{\partial \bar{\theta}^{\dot{\alpha}}} \frac{\partial}{\partial \bar{\theta}_{\dot{\alpha}}}e^{-\hat{V}_i}\frac{\partial}{\partial \theta^\alpha} e^{\hat{V}_i}.
\end{align}
We now show that this reduced model reproduces Eq.(\ref{action1}) in bosonic non-commutative space.
The equation of motion of Eq.(\ref{reduced}) for $\hat{V}_i$ is given by
\begin{align}
\frac{\partial}{\partial \theta^\alpha } e^{\hat{V}_i} \hat{W}_i^\alpha e^{-\hat{V}_i}=0,
\end{align}
which has the special solution
\begin{align}
\exp\hat{V}_i(\theta,\bar{\theta})=
\exp\left( -2 \theta \sigma^\mu \bar{\theta} \hat{p}_\mu \right)\otimes 1_{N_i},
\label{solution}
\end{align}
where $\hat{p}_\mu$ is given by Eq.(\ref{bosm2}).
We can expand $e^{\hat{V}_i}$ around this background as
\begin{align}
e^{\hat{V}_i}&=e^{\hat{A}}e^{\hat{V'}_i}e^{\hat{A}}, \nonumber \\
\hat{A}&\equiv -\theta \sigma^\mu \bar{\theta}  \hat{p}_\mu \otimes 1_{N_i},
\end{align}
Action $S$ then becomes
\begin{align}
S=& 
(2\pi)^2 \sqrt{\det C}\int  d^2 \theta d^2 \bar{\theta}~\Biggl\{ \sum_{i=1}^n  \Tr_{U(\hat{N}_i)} ( \hat{\bar{\Phi'}}_i e^{\hat{V'}_i} \hat{\Phi'}_{i} e^{-\hat{V'}_i}) \nonumber \\
&+\sum_{i>j} \Tr_{U(\hat{N}_j)}(\hat{\bar{X'}}_{i,j} e^{\hat{V'}_i} \hat{X'}_{i,j} e^{-\hat{V'}_j}) 
+\sum_{i>j} \Tr_{U(\hat{N}_j)}(\hat{\bar{Y'}}_{j,i} e^{\hat{V'}_j} \hat{Y'}_{j,i} e^{-\hat{V'}_i}) \Biggr\} \nonumber \\
&+ (2\pi)^2 \sqrt{\det C} \int d^2 \theta~ \Biggl\{ \sum_{i=1}^n 2\pi i \tau_i \Tr_{U(\hat{N}_i)}\hat{W'}_i^\alpha \hat{W'}_{i\alpha} \nonumber \\
&+ \sum_{(i,a)}g_{(i,a)}\Tr_{U(\hat{N}_i)}O_{(i,a)}\left( \hat{\Phi}'(\theta), \hat{X}'(\theta),\hat{Y}'(\theta) \right)  \Bigr\}+ c.c..
\label{reduced2}
\end{align}
Here, 
\begin{align}
\hat{\Phi'}_i(\theta,\bar{\theta}) & = e^{\hat{A}} \hat{\Phi}_i(\theta) e^{-\hat{A}} , \nonumber \\
\hat{X'}_{ij}(\theta,\bar{\theta}) & = e^{\hat{A}} \hat{X}_{ij}(\theta) e^{-\hat{A}} , \nonumber \\
\hat{Y'}_{ij}(\theta,\bar{\theta}) & = e^{\hat{A}} \hat{Y}_{ij}(\theta) e^{-\hat{A}} , \nonumber \\
\hat{\bar{\Phi'}}_i(\theta,\bar{\theta}) &=e^{-\hat{A}} \hat{\bar{\Phi}}_i(\bar{\theta}) e^{\hat{A}} , \nonumber \\
{\hat{W}'}_{i\alpha}(\theta,\bar{\theta})&=e^{\hat{A}}\hat{W}_{i\alpha}(\theta)e^{-\hat{A}}=-\frac{1}{4}\hat{\bar{D}}\hat{\bar{D}}e^{-{\hat{V}'}_i}\hat{D}_\alpha e^{{\hat{V}'}_i}, \nonumber \\
\hat{D}_\alpha&=e^{-\hat{A}}\frac{\partial}{\partial \theta^\alpha}e^{\hat{A}}=
\frac{\partial}{\partial \theta^\alpha}-\left(\sigma^\mu \bar{\theta}\right) \hat{p}_\mu, \nonumber \\
{\hat{\bar{D}}}_{\dot{\alpha}}
&= e^{\hat{A}} 
\left(-\frac{\partial}{ \partial {\bar{\theta}}^{\dot{\alpha}} } \right)
e^{-\hat{A}} 
= -\frac{\partial}{\partial {\bar{\theta}}^{\dot{\alpha}}}
+ \left( \theta \sigma^\mu \right)  {\hat{p}}_\mu.
\end{align}
We can now map these matrices to superfields through Weyl ordering (Eq.(\ref{Weyl2})) as follows:
\begin{align}
\hat{\Phi'}_i(\theta,\bar{\theta}) & \mapsto \Phi_i(x,\theta,\bar{\theta}) \nonumber \\
\hat{X'}_{ij}(\theta,\bar{\theta}) & \mapsto X_{ij}(x,\theta,\bar{\theta}) \nonumber \\
\hat{Y'}_{ij}(\theta,\bar{\theta}) & \mapsto Y_{ij}(x,\theta,\bar{\theta}) \nonumber \\
\hat{\bar{\Phi'}}_i(\theta,\bar{\theta}) & \mapsto \bar{\Phi}_i(x,\theta,\bar{\theta}) , \nonumber \\
\hat{V'}_i(\theta,\bar{\theta}) & \mapsto V_i(x,\theta,\bar{\theta}) \nonumber \\
{\hat{W'}}_{i\alpha}(\theta,\bar{\theta})& \mapsto W_{i\alpha}(x,\theta,\bar{\theta}). \nonumber 
\end{align}
We can also map the action in Eq.(\ref{reduced2}) to non-commutative field theory as in Eq.(\ref{map10}). 
As a result, this reduced model is equivalent to a non-commutative action which can be obtained by replacing the product in Eq.(\ref{action1}) with star products in Eq.(\ref{star1}).  

Note that the ratios of the infinite ranks of the gauge groups are fixed in order to reproduce Eq.(\ref{action1}).
It is obvious, if we define the infinite rank of $\hat{p}_\mu$ as $\hat{M}$ in Eq.(\ref{solution}), we can obtain the relationships
\begin{align}
\frac{\hat{N}_i}{\hat{N}_j}=\frac{\hat{M}\times N_i}{\hat{M}\times N_j}=\frac{N_i}{N_j}.
\label{ratio}
\end{align}

\subsection{Construction of the supermatrix model}
\label{2.4}
We consider Eq.(\ref{reduced}) in the fermionic non-commutative space \cite{NCS} (Eq.(\ref{ferc})) by replacing the product with the star products in Eq.(\ref{star2}) as follows: 
\begin{align}
S=& 
 (2\pi)^2 \sqrt{\det C}\int  d^2 \theta d^2 \bar{\theta}~\Biggl\{ \sum_{i=1}^n \Tr_{U(\hat{N}_i)} ( \hat{\bar{\Phi}}_i e^{\hat{V}_i} \hat{\Phi}_{i} e^{-\hat{V}_i}) \nonumber \\
&+\sum_{i>j} \Tr_{U(\hat{N}_j)}(\hat{\bar{X}}_{i,j} e^{\hat{V}_i} \hat{X}_{i,j} e^{-\hat{V}_j}) 
+\sum_{i>j} \Tr_{U(\hat{N}_j)}(\hat{\bar{Y}}_{j,i} e^{\hat{V}_j} \hat{Y}_{j,i} e^{-\hat{V}_i}) \Biggr\}_\star \nonumber \\
&+ (2\pi)^2 \sqrt{\det C} \int d^2 \theta~ \Biggl\{ \sum_{i=1}^n 2\pi i \tau_i \Tr_{U(\hat{N}_i)}\hat{W}_i^\alpha \hat{W}_{i\alpha} \nonumber \\
&+ \sum_{(i,a)}g_{(i,a)}\Tr_{U(\hat{N}_i)}O_{(i,a)}\left( \hat{\Phi}(\theta), \hat{X}(\theta),\hat{Y}(\theta) \right)  \Biggr\}_\star+ c.c..
\label{Sreduced}
\end{align}
We can then map this action to the supermatrix model as shown in Appendix \ref{a2} and obtain 
\begin{align}
S=& 
\frac{i^2(2\pi)^2\sqrt{\det C}}{8^2 \sqrt{\det \gamma}\sqrt{\det \gamma^*}} 
\Bigl\{ \sum_{i=1}^n \Str_{U(\hat{N}_i)} ( \hat{\bar{\Phi}}_i e^{\hat{V}_i} \hat{\Phi}_{i} e^{-\hat{V}_i}) \nonumber \\
&+\sum_{i>j} \Str_{U(\hat{N}_j)}(\hat{\bar{X}}_{i,j} e^{\hat{V}_i} \hat{X}_{i,j} e^{-\hat{V}_j}) 
+\sum_{i>j} \Str_{U(\hat{N}_j)}(\hat{\bar{Y}}_{j,i} e^{\hat{V}_j} \hat{Y}_{j,i} e^{-\hat{V}_i}) \Bigr\} \nonumber \\
&+ \frac{i(2\pi)^2\sqrt{\det C}}{8 \sqrt{\det \gamma}} \Bigl\{ 
2\pi i \tau_i \Str_{U(\hat{N}_i)}\left( \hat{W}_i^{\alpha} \hat{W}_{i\alpha} \right) \nonumber \\
&+ \sum_{(i,a)}g_{(i,a)}\Str_{U(\hat{N}_i)}O_{(i,a)}\left( \hat{\Phi}, \hat{X},\hat{Y} \right)  \Bigr\}
+ c.c..
\label{smm2}
\end{align}
The matrices here can be obtained by Weyl ordering (Eq.(\ref{Weyl})).
In particular, the field strength is
\begin{align}
\hat{W}_{i\alpha}=-\frac{1}{4} \ad \hat{\bar{\pi}}_{\dot{\alpha}} \ad \hat{\bar{\pi}}^{\dot{\alpha}}e^{-\hat{V}_i}\ad \hat{\pi}_\alpha e^{\hat{V}_i}.
\end{align}
In precise terms, $\Str_{U(\hat{N}_i)}$ means $\Str_{U(\hat{N}_i|\hat{N}_i)}$ in  (anti-) chiral terms or $\Str_{U(\hat{N}_i|\hat{N}_i )\otimes U(1|1)}$ in matter kinetic terms
 (the definition of the $\Str$ is given in Eq.(\ref{Str}).).\\

Let us now consider the relationships to non-commutative field theory.
The action in Eq.(\ref{reduced}) does not depend on $\hat{p}_\mu$, and this dependence appears when we consider the fluctuation around the classical solution of the gauge fields (Eq.(\ref{solution})).
We can then map the action to non-commutative field theory.
However, it is not obvious that the action in Eq.(\ref{smm2}) has an appropriate classical solution for $\hat{V}_i$, as in the case for Eq.(\ref{solution}) \cite{ST}.  
Even if we find such a solution, it is also not obvious that the fluctuation around the solution can be regarded as a non-commutative field.
However, this problem is not serious when we take the limit $\gamma, \gamma^* \rightarrow0$.\\

We next consider the commutative limit of Eq.(\ref{Sreduced}) and (\ref{smm2}).
In order to obtain the commutative field theory (Eq.(\ref{action1})) from Eq.(\ref{Sreduced}),
we first take the limit of $\gamma, \gamma^* \rightarrow 0$. 
We may then obtain Eq.(\ref{reduced}) and this action can be mapped to the action in Eq.(\ref{action1}) in bosonic non-commutative coordinates by considering the fluctuation around Eq.(\ref{solution}).
Lastly, we take the limit of $C \rightarrow 0$ and obtain the commutative field theory (Eq.(\ref{action1})).
\begin{align*}
\text{Supermatrix model (\ref{smm2})}
&=~\text{(\ref{Sreduced})} \\
&\xrightarrow{\gamma,\gamma^* \rightarrow 0} \text{(\ref{reduced})} \\
&= ~\text{(\ref{action1}) in bosonic non-commutative coordinates} \\
&\xrightarrow{C\rightarrow 0} \text{(\ref{action1})}
\end{align*}
We take the two commutative limits in these processes.
The first limit ($\gamma \rightarrow 0$) can be taken smoothly owing to the non-commutativity $C$. 
On the other hand, under the second limit ($C \rightarrow 0$), non-commutative field theories may not correspond to the commutative field theories (UV/IR mixing).
In this sense, it is not clear that the commutative limits of the supermatrix model correspond to the commutative field theory.

However, the holomorphic parts of the supermatrix model almost correspond to the commutative field theory. 
Under the second limit,
the chiral ring property eliminates the non-planar diagrams which give the non-commutative phase to the amplitude.
The difference only appears through the Konishi anomaly \cite{Konishi}.
This anomaly in commutative field theory appears through the regularization procedure.
On the other hand, 
in non-commutative field theory, such regularization is not necessary, since the non-commutativity $C$ 
regularizes such singularity (see Sec. \ref{2.5}).\\

In the case of Eq.(\ref{smm2}),
the holomorphic parts correspond to the commutative field theory.
As a result of this holomorphy, the matter kinetic terms and anti-holomorphic parts can be neglected when we consider the holomorphic quantities.
Dropping the matter kinetic terms implies that the matter is decoupled from the gauge fields and that we can also neglect the gauge kinetic terms. 
We can thus obtain the following supermatrix model:
\begin{align}
S_{smm}&=\frac{\hat{N}}{g_m} \sum_{(i,a)}g_{(i,a)}  \Str_{U(\hat{N}_i)}O_{(i,a)}\left( \hat{\Phi}, \hat{X},\hat{Y} \right) 
\label{SMM}
\end{align}
where $\hat{N}/g_m$ is given in Eq.(\ref{coefficient}) as follows:
\begin{align*}
\frac{\hat{N}}{g_m}=\frac{i(2\pi)^2\sqrt{\det C}}{8\sqrt{\det \gamma}}.
\end{align*}
The dependence on non-commutativity $C,\gamma$ in this model only appears through the overall factor, $\hat{N}/g_m$.
As a result of this structure, the loop equations of this model have no dependence on $C,\gamma$ 
(we show this example in section \ref{3.3}).
This means that, by taking $C,\gamma \rightarrow 0$, this supermatrix model has a commutative limit.
Some of the physical quantities which can be calculated by this supermatrix model  correspond to the quantities by the field theory as shown in subsequent subsections.
This action does not depend on the gauge fields.
However, when we map the correlation function of the supermatrix model to non-commutative field theory under $C,\gamma \rightarrow 0$,
 singularity appears in the field theory.
Upon regularizing such singularity, the dependence on the gauge fields appears in the field theory.

The result enables us to conclude that the supermatrix model (Eq.(\ref{SMM}))  corresponds to the holomorphic part of the commutative field theory (\ref{action1}) in the limit of $C,\gamma \rightarrow 0$.

This implies that the commutative limit of Eq.(\ref{Sreduced}) corresponds to the commutative field theory.

\subsection{Correspondence of the correlation functions}
\label{2.5}
We will show that
\begin{align}
\frac{1}{64\pi^2}\left< \tr_{U(N_i)} W_i^\alpha W_{i\alpha}  P(\Phi,X,Y) \right>
= \frac{g_m}{\hat{N}}\left< \Str_{U(\hat{N_i})}  P(\hat{\Phi},\hat{X},\hat{Y}) \right>,
\label{proof1}
\end{align}
in the limit, $C,\gamma \rightarrow 0$, where $P(\Phi,X,Y)$ is a polynomial function and has  $U(N_i)$  gauge indices.
The left-hand side of this equation is a gauge theory correlation function, and the right-hand side is a supermatrix model function.
This provides evidence that our supermatrix model includes the information of the field theory.

\paragraph{Proof}
We use the identity
\begin{align}
\left(\delta^4(\hat{x}-x) \delta^2(\hat{\theta}-\theta)\right)^2=\frac{g_m^2}{\hat{N}^2}
=-\frac{64\det{\gamma}}{(2\pi)^4\det{C}}.
\end{align}
This identity holds in non-commutative space \cite{KKM}.
Inserting this on the right-hand side of Eq.(\ref{proof1}), we obtain
\begin{align*}
&\frac{g_m}{\hat{N}}
\left< \Str_{U(\hat{N}_i)} \left(  P\left(\hat{\Phi},\hat{X},\hat{Y}\right)  \right) \right> \\
&=\frac{i(2\pi)^2\sqrt{\det C}}{8\sqrt{\det \gamma}}
\left< \Str_{U(\hat{N}_i)} \left( \left(\delta^4(\hat{x}-x) \delta^2(\hat{\theta}-\theta)\right)^2  P\left(\hat{\Phi},\hat{X},\hat{Y}\right)  \right) \right>
\end{align*}
We can map this supermatrix correlator to the fermionic non-commutative field theory as follows: 
\begin{align*}
(2\pi)^2\sqrt{\det C}
\left< \int d^2 \theta'~ \Tr_{U(\hat{N}_i)} \left( \left(\delta^4(\hat{x}-x) \delta^2(\theta'-\theta)\right)^2  P\left(\hat{\Phi}(\theta),\hat{X}(\theta),\hat{Y}(\theta)\right)  \right) \right>_{\star},
\end{align*}
where $\star$ indicates that $\theta$ is non-commutative. By taking $\gamma \rightarrow 0$, we obtain the matrix correlator,
\begin{align*}
(2\pi)^2\sqrt{\det C}
\left< \int d^2 \theta'~\Tr_{U(\hat{N}_i)} \left( \left(\delta^4(\hat{x}-x) \delta^2(\theta'-\theta)\right)^2  P\left(\hat{\Phi}(\theta),\hat{X}(\theta),\hat{Y}(\theta)\right)  \right) \right>.
\end{align*}
This correlator can be mapped to one for bosonic non-commutative field theory by considering the fluctuation around the classical solution in Eq.(\ref{solution}) as follows:
\begin{align*}
\left< \int d^4x' d^2\theta'~\tr_{U(N_i)} \left( \left(\delta^4(x'-x) \delta^2(\theta'-\theta)\right)^2  P \left(\Phi(x',\theta'),X(x',\theta'),Y(x',\theta')\right)  \right) \right>_* ,
\end{align*}
and by taking $C \rightarrow 0$, we obtain 
\begin{align}
\left< \int d^4x' d^2\theta' ~\tr_{U(N_i)}\left( \left(\delta^4(x'-x) \delta^2(\theta'-\theta)\right)^2  P \left(\Phi(x',\theta'),X(x',\theta'),Y(x',\theta')\right)  \right) \right>.
\label{eq1}
\end{align}
When we integrate $x'$ and $\theta'$, 
 singular factor, $\delta^4(0)\delta^2(0)$, appears and we can regularize this by the heat kernel method shown in the appendix to ref.\cite{KKM} as follows:
\begin{align}
\left. \delta^4(x'-x)\delta^2(\theta'-\theta){\delta^k}_j \right|_{(x',\theta')\mapsto(x,\theta)}
=\frac{1}{64\pi^2}{\left(W_i^\alpha  W_{i\alpha} \right)^k}_j
\label{Konishi}
\end{align}
where $j,k$ are indices for the representation of the $U(N_i)$ gauge group.
The right-hand side of this equation is the contribution from the Konishi anomaly \cite{Konishi} of the $U(N_i)$ gauge field.

Eq.(\ref{eq1}) therefore becomes 
\begin{align}
\left<\frac{1}{64\pi^2} \tr_{U(N_i)} W_i^\alpha W_{i\alpha}  P(\Phi,X,Y) \right>.
\end{align}
This is the left-hand side of Eq.(\ref{proof1}).

On the other hand, the correlators 
\begin{align*}
\left< \frac{g_m}{\hat{N}} \Str_{U(\hat{N}_i)} P(\hat{\Phi},\hat{X},\hat{Y}) \right>
\end{align*}
are independent of $g_m/\hat{N}$ (or $C$ and $\gamma$) without an overall factor.
This can be explicitly checked when we solve the loop equation of these correlators \cite{KKM}.

This foregoing analysis indicates that, when we take $C,\gamma \rightarrow 0$, we can obtain
\begin{align}
\left<\frac{1}{64\pi^2}
\tr_{U(N_i)} W^\alpha_i W_{i\alpha} P(\Phi,X,Y) \right>
=\left< \frac{g_m}{\hat{N}} \Str_{U(\hat{N}_i)} P(\hat{\Phi},\hat{X},\hat{Y}) \right>.
\end{align}

\subsection{Low-energy effective action and prepotential}
\label{2.6}
In this subsection, we consider the property of the effective action of Eq.(\ref{action1}) and  show that it can be described by a prepotential.

The coupling $g_{(i,a)}$ dependence of $W_{eff}$ is obtained by differentiating the partition function with respect to $g_{(i,a)}$ as follows: 
\begin{align}
\frac{\partial}{\partial g_{(i,a)}}W_{eff}
=& \left< \tr_{U(N_i)}O_{(i,a)}(\Phi,X,Y) \right>, \nonumber 
\end{align}
We can express the right-hand side as
\begin{align}
\int d^2 \psi \frac{1}{64\pi^2} \left< \tr_{U(N_i)}(W^\alpha_i -8\pi\psi^\alpha 1_{N_i})(W_{i\alpha}-8\pi\psi_\alpha 1_{N_i})O_{(i,a)}(\Phi,X,Y)\right>,
\label{eff}
\end{align}
where $\psi$ is an anti-commuting c-number.

Let us now consider the special case that the loop equation for 
\begin{align}
\left< \tr_{U(N_i)} W_i^\alpha W_{i\alpha} O_{(i,a)}(\Phi,X,Y)\right>
\end{align}
is closed under single-trace correlators which include just two field strengths:
\begin{align}
\left< \tr \left( W_j^\alpha W_{j\alpha}\cdots \right) \right>.
\label{operator}
\end{align}
We can then obtain a new loop equation for
 \begin{align}
 \left< \tr(W^\alpha_i -8\pi\psi^\alpha)(W_{i\alpha}-8\pi\psi_\alpha ) O_{(i,a)}(\Phi,X,Y) \right>
 \label{operator2}
 \end{align}
through the shift 
\begin{align}
W_{i\alpha} \mapsto W_{i\alpha} - 8\pi \psi_{\alpha}1_{N_i},
\label{shift}
\end{align} 
for all $i$.
Under this shift, the correlators do not change, 
since the action in Eq.(\ref{action1}) has symmetry under this shift.
\footnote{
If we rewrite the kinetic terms of the action (\ref{action1}) as
\begin{align*}
\int d^8z~ \tr_{U(N)} \bar{\Omega} e^{\Xi} \Omega e^{-\Xi}, 
\end{align*}
 where $N=\sum_{i=1}^n N_i$ and
 \begin{align}
\Omega &=
\begin{pmatrix}
\Phi_1 & Y_{1,2} & \ldots & \\
X_{2,1} & \Phi_2 & & \\
\vdots &       & \ddots &  \\
 & & & \Phi_n
 \end{pmatrix},\\
 \Xi&=
 \begin{pmatrix}
 V_1 & 0  & \ldots & \\
0 & V_2 &  & \\
\vdots &       & \ddots &  \\
 & & & V_n
 \end{pmatrix},
 \end{align}
then the symmetry (\ref{shift}) is regarded as that of $U(N)$ gauge and one adjoint matter case.
Thus this symmetry is obvious.
}
This loop equation is closed with respect to the operators
\begin{align*}
\left< \tr_{U(N_i)}(W^\alpha_i -8\pi\psi^\alpha)(W_{i\alpha}-8\pi\psi_\alpha) \cdots\right>,
\end{align*}
which are obtained by shifting Eq.(\ref{operator}).
These operators are related to each other through the loop equations, 
and we can pick up independent operators from among them.  
(In the case of the $U(N)$ gauge theory coupled to a single adjoint matter, such independent operators are glueball superfields $\mathcal{S}_i$ (not $S_i$) in ref.\cite{CDSW}.)
$W_{eff}$ can then be expressed by prepotential $\mathcal{F}$, which is a function of these independent operators, as
\begin{align}
W_{eff}=\int d^2\psi \mathcal{F}.
\label{pre}
\end{align}
Comparing this and Eq.(\ref{eff}), we can obtain the following equation for $\mathcal{F}$:
\begin{align}
\frac{\partial}{\partial g_{(i,a)}} \mathcal{F}
=  \frac{1}{64\pi^2} \left< \tr_{U(N_i)}(W^\alpha_i -8\pi\psi^\alpha)(W_{i\alpha}-8\pi\psi_\alpha)O_{(i,a)}(\Phi,X,Y)\right>.
\label{prepotential}
\end{align}
In order to compute the prepotential, we solve the loop equations for the correlator on the right-hand side with respect to the independent operators.
However, the purpose of this study is to show that the correlator in Eq.(\ref{prepotential}) is equivalent to a certain correlator in the supermatrix model 
by direct mapping, instead of by solving the loop equations.

\subsection{Free energy of the supermatrix model and prepotential}
\label{2.3}
We evaluate in this section the free energy of the supermatrix model 
and consider its relationship to the prepotential.

The free energy of the supermatrix model is defined by
\begin{align}
\exp\left( -\frac{\hat{N}^2}{g_m^2} F_m \right) 
= \int d \hat{\Phi} d \hat{X} d \hat{Y}~e^{-S_{smm}}.
\end{align}
In order to find the dependence on $g_{(i,a)}$, we differentiate this equation with respect to $g_{(i,a)}$ as follows:
\begin{align}
\frac{\partial}{\partial g_{(i,a)}}F_m 
=  \frac{g_m}{\hat{N}}\left< \Str_{U(\hat{N}_i)} O_{(i,a)}(\hat{\Phi},\hat{X},\hat{Y}) \right>.
\end{align}
If the  correlation function on the right-hand side is equivalent to Eq.(\ref{prepotential}), the free energy and prepotential are equivalent up to the $g_{(i,a)}$-independent part.

We can now show
\begin{align}
\left<\frac{1}{64\pi^2}
\tr_{U(N_i)}(W^\alpha_i-8\pi\psi^{\alpha})( W_{i\alpha}-8\pi\psi_{\alpha}) O_{(i,a)}(\Phi,X,Y) \right>
=\frac{g_m}{\hat{N}}\left< \Str_{U(\hat{N}_i)} O_{(i,a)}(\hat{\Phi},\hat{X},\hat{Y}) \right>
\label{equivalence}
\end{align}
under the commutative limit.
As in subsection \ref{2.5}, the right-hand side is equivalent to
\begin{align*}
\left<\frac{1}{64\pi^2}
\tr_{U(N_i)}W^\alpha_i W_{i\alpha} O_{(i,a)}(\Phi,X,Y) \right>.
\end{align*}
Due to the symmetry in  Eq.(\ref{shift}), this becomes
\begin{align*}
\left<\frac{1}{64\pi^2}
\tr_{U(N_i)}(W^\alpha_i-8\pi\psi^{\alpha})( W_{i\alpha}-8\pi\psi_{\alpha}) O_{(i,a)}(\Phi,X,Y) \right>.
\end{align*}
This is the left-hand side of Eq.(\ref{equivalence}).
This equivalence means that the free energy of the supermatrix model is equivalent to the prepotential up to the $g_{(i,a)}$ -independent part.

The equivalence of the coupling constant of the independent parts can be checked
in a discrete calculation with a simple form of the superpotential;- for example, gaussian or cubic.
The result can be generalized to any superpotential, since such terms do not depend on the form of the superpotential.
However, with our approach, we have not yet provided a good explanation for why the reduced model  reproduces the $g_{(i,a)}$-independent part of the prepotential as well. \\

Note that the symmetry in Eq.(\ref{shift}) means that the gauge field correlation functions 
\begin{align*}
 \frac{1}{64\pi^2}\left< \tr W^\alpha W_\alpha \cdots \right>
 ,~~\frac{1}{8\pi}\left< \tr W_\alpha \cdots \right>
 ,~~\left< \tr \cdots \right>
\end{align*}
behave as one chiral multiplet,
\begin{align*}
\frac{1}{64\pi^2}\left< \tr(W^\alpha-8\pi\psi^\alpha)( W_\alpha-8\pi\psi_\alpha) \cdots \right>. 
\end{align*}
The equivalence in Eq.(\ref{equivalence}) means that this multiplet corresponds to the supermatrix correlation function 
\begin{align*}
\frac{g_m}{\hat{N}}\left<\Str \cdots \right>.
\end{align*}
As a result, all the chiral operators of the gauge field corresponds to the supermatrix correlators   through this multiplet.\\

Our supermatrix model can therefore describe the holomorphic part of the gauge theory.
In this respect, the Dijkgraaf-Vafa theory with multiple bifundamental matter can be regarded as large-$N$ reduction
(the difference between the supermatrix and bosonic matrix is considered in ref.\cite{KKM}).

\section{Supersymmetric reduced model with fundamental matters}
\label{3}
\setcounter{equation}{0}
We consider in this section the $A_2$ quiver-like theory,
from which we can develop a theory which incorporates fundamental matter. 

\subsection{Construction of the theory with fundamental matters}
We will evaluate a theory which has gauge groups (Eq.(\ref{A_2})) and  $A_2$ quiver-like kinetic terms.
The action is given by
\begin{align} 
S=&\int d^4xd^2\theta d^2\bar{\theta}~\left( \tr_{U(n_c)}(\bar{\Phi}e^V \Phi e^{-V})
+ \tr_{U(n_f)}(\bar{Q}e^V Q e^{-V_f})  
+ \tr_{U(n_f)}( \bar{\tilde{Q}} e^{V_f} \tilde{Q} e^{-V} ) \right) \nonumber \\
+ &\int d^4x d^2\theta \left( ~2\pi i\tau~\tr_{U(n_c)}(W^\alpha W_\alpha)
+ 2\pi i\tau_f~\tr_{U(n_f)}(W_f^\alpha W_{f\alpha}) \right) \nonumber \\
+ &\int d^4x d^2\theta ~\left( \tr_{U(n_c)}(W(\Phi))+ \tr_{U(n_f)}(\tilde{Q}m(\Phi) Q)\right)
+ c.c..
\label{action2}
\end{align}
$V$ is a vector superfield of gauge group $U(n_c)$, $W_\alpha$ is its field strength, $V_f$ and $W_{f\alpha}$ are the vector superfield and field strength of the $U(n_f)$, $Q$ and $\tilde{Q}$ are (anti)bifundamental matter and $\Phi$ is adjoint matter of $U(n_c)$,
and $W(\Phi)$ and $m(\Phi)$ are polynomial functions.
The adjoint matter of $U(n_f)$ in this theory is removed  from the usual $A_2$ theory.\footnote{
Or we give  an adjoint matter $\Phi_{n_f}$ for $U(n_f)$ an infinite mass :$\frac{m}{2}\tr \Phi_{n_f}^2 $. 
Then $\Phi_{n_f}$ is decoupled. }

When we turn off gauge field $V_f$ and regard the $U(n_f)$ symmetry as a global,
bifundamental matter $Q$ behaves as a fundamental matter with $U(n_f)$ flavor symmetry.
Since we can construct a supermatrix model in the general gauge theory with bifundamental matter as described in section \ref{2}, we first keep this $U(n_f)$ symmetry as gauge symmetry and ultimately take $V_f=0$. 

The holomorphic parts of the supermatrix model for the action Eq.(\ref{action2}) become
\begin{align}
S_{smm}=\frac{\hat{N}}{g_m}\left( \Str_{U(\hat{n}_c)}\left( W(\hat{\Phi}) \right)
+ \Str_{U(\hat{n}_f)} \left( \hat{\tilde{Q}}m(\hat{\Phi}) \hat{Q} \right) \right),
\label{action3}
\end{align}
where $\hat{Q}$, $\hat{\tilde{Q}}$ and $\hat{\Phi}$ are matrices corresponding to $Q$, $\tilde{Q}$ and $\Phi$.
This supermatrix model has no gauge field dependence.
When we take $V_f=0$, the supermatrix action does not change.

Note that this supermatrix model is in  the Veneziano limit \cite{Ven} in which the number of colors and flavors become infinite with their ratio fixed as in Eq.(\ref{ratio}).
 Even if we take  $V_f$ to be $0$, there is no influence on the size of the matrices. 

\subsection{Derivation of the loop equation from the supermatrix model}
\label{3.3}
We derive in this subsection the loop equations from the supermatrix model, and show that these equations are closed and that the dependence on $C,\gamma$ disappears.
Furthermore, these loop equations can be mapped to those of the field theory, which is 
 equivalent to the field theory analyses \cite{S}, when we take $V_f=0$.
This result justifies our supermatrix approach in section \ref{2}.\\

We start from
\begin{align*}
\left< \frac{g_m}{\hat{N}}\Str_{U(\hat{n}_c)}\left(T^a \frac{1}{z-\hat{\Phi}}\right) \right>,
\end{align*}
where $T^a$ is the Gell-Mann matrix.
We shift $\hat{\Phi} \mapsto \hat{\Phi} + \epsilon T^a$ and obtain the Schwinger-Dyson equation,
\begin{align}
0=&\frac{g_m}{\hat{N}}\left< \Str_{U(\hat{n}_c)}\left( T^a\frac{1}{z-\hat{\Phi}}T^a\frac{1}{z-\hat{\Phi}}\right) \right> 
-\left< \Str_{U(\hat{n}_c)}\left(T^a \frac{1}{z-\hat{\Phi}}\right) \Str_{U(\hat{n}_c)}\left(T^a W'(\hat{\Phi}) \right) \right> \nonumber \\
&-\left<\Str_{U(\hat{n}_c)}\left(T^a \frac{1}{z-\hat{\Phi}}\right) \Str_{U(\hat{n}_f)}\left( \hat{\tilde{Q}} T^a m'(\hat{\Phi})\hat{Q} \right) \right>.
\end{align}
By using the completeness of the Gell-Mann matrix and large-$N$ factorization, we can obtain the loop equation,
\begin{align}
\left(  \frac{g_m}{\hat{N}}\left< \Str_{U(\hat{n}_c)}\frac{1}{z-\hat{\Phi}}\right> \right)^2 
- \frac{g_m}{\hat{N}} \left< \Str_{U(\hat{n}_c)}\frac{W'(\hat{\Phi})}{z-\hat{\Phi}} \right>
- \frac{g_m}{\hat{N}} \left< \Str_{U(\hat{n}_f)}\hat{\tilde{Q}}\frac{m'(\hat{\Phi})}{z-\hat{\Phi}}\hat{Q}\right>=0.
\label{smmloop}
\end{align}

We next start from
\begin{align}
\left< \Str_{U(\hat{n}_f)}\hat{\tilde{Q}} \frac{1}{z-\hat{\Phi}}T^a \right>.
\end{align}
By shifting $\hat{\tilde{Q}}\mapsto \hat{\tilde{Q}}+ \epsilon T^a$, we can derive another loop equation.
\begin{align}
 \frac{g_m}{\hat{N}} \left< \Str_{U(\hat{n}_f)}1 \right>\frac{g_m}{\hat{N}} \left<\Str_{U(\hat{n}_c)}\frac{1}{z-\hat{\Phi}} \right>
= \left< \frac{g_m}{\hat{N}}\Str_{U(\hat{n}_f)} \hat{\tilde{Q}} \frac{m(\hat{\Phi})}{z-\hat{\Phi}}\hat{Q} \right>.
\label{smmloop2}
\end{align}

As already mentioned, the dependence on $C,\gamma$ in these loop equations appears only through $g_m/\hat{N}$.
These two loop equations are closed and can be solved as a function of
\begin{align*}
S_i=\int_{C_i}dz~\frac{g_m}{\hat{N}}\left< \Str \frac{1}{z-\hat{\Phi}}\right>,
\end{align*}
where $C_i$ is a contour around the $i$-th critical point of superpotential $W(\hat{\Phi})$.
The $g_m/\hat{N}$ dependence disappears if we solve the correlators with respect to $S_i$.
These loop equations are thus independent of $C,\gamma$ and have a commutative limit.\\

We will next show that this commutative limit corresponds to the loop equations which are derived from gauge theory.
In order to map these loop equations to the gauge theory equations, we repeat the same procedure as that for deriving Eq.(\ref{equivalence}).
This enables us to obtain the loop equations
\begin{align}
&\left( \frac{1}{64\pi^2}\left<  \tr_{U(n_c)}\frac{(W^\alpha- 8\pi\psi^\alpha)( W_\alpha -8\pi\psi_\alpha)}{z-\Phi^{(y)}(x,\theta)}\right> \right)^2 \nonumber \\
&-  \frac{1}{64\pi^2}\left< \tr_{U(n_c)}\frac{(W^\alpha-8\pi\psi^\alpha)( W_\alpha-8\pi\psi_\alpha) W'(\Phi^{(y)}(x,\theta))}{z-\Phi^{(y)}(x,\theta)} \right> \nonumber \\
&- \frac{1}{64\pi^2} \left<\tr_{U(n_f)}(W^\alpha_f-8\pi\psi^\alpha)( W_{f\alpha} -8\pi\psi_\alpha)  \tilde{Q}^{(y)}\frac{m'(\Phi^{(y)})}{z-\Phi^{(y)}}Q^{(y)}(x,\theta)\right>=0,\\
&\frac{1}{64\pi^2} \left< \tr_{U(n_f)} \left( (W^\alpha_f-8\pi\psi^\alpha)(W_{f\alpha} -8\pi\psi_\alpha)\right) \right>\left<  \frac{1}{64\pi^2} \tr_{U(n_c)} \left(\frac{W^\alpha W_\alpha}{z-\Phi}\right) \right>
\nonumber \\
&= \frac{1}{64\pi^2} \left< \tr_{U(n_f)}\left((W^\alpha_f-8\pi\psi^\alpha)(W_{f\alpha} -8\pi\psi_\alpha) \tilde{Q}\frac{m(\Phi)}{z-\Phi}Q\right)\right>.
\label{loop5}
\end{align}
These equations are closed with respect to the following correlators: 
\begin{align*}
\left< \tr\left( (W^\alpha- 8\pi\psi^\alpha)( W_\alpha -8\pi\psi_\alpha) \cdots \right) \right>, \\
\left< \tr\left( (W^\alpha_f-8\pi\psi^\alpha)(W_{f\alpha} -8\pi\psi_\alpha) \cdots \right) \right>.
\end{align*}
This fact leads to an efficient description of the prepotential as mentioned in section \ref{2.3}.

Expanding these equations with respect to $\psi_\alpha$ and turning off the formal gauge field strength, $W_{f\alpha}$, we can obtain the following loop equations:
\begin{align}
&\left< \tr_{U(n_c)}\frac{W'(\Phi)}{z-\Phi} \right> + \left< \tr_{U(n_f)}\tilde{Q}\frac{m'(\Phi)}{z-\Phi}Q \right>
= \nonumber \\
&2\frac{1}{64\pi^2} \left< \tr_{U(n_c)}\frac{W^\alpha W_\alpha}{z-\Phi} \right>\left< \tr_{U(n_c)}\frac{1}{z-\Phi}\right> 
+\frac{1}{8\pi}\left< \tr_{U(n_c)}\frac{W^\alpha}{z-\Phi} \right>\frac{1}{8\pi} \left<\tr_{U(n_c)}\frac{W_\alpha}{z-\Phi}\right>,
\label{loop1}\\
&\frac{1}{8\pi}\left< \tr_{U(n_c)}\frac{W'(\Phi)W_\alpha}{z-\Phi}\right>
=2\frac{1}{64\pi^2} \left< \tr_{U(n_c)}\frac{W^\alpha W_\alpha}{z-\Phi}\right>\frac{1}{8\pi} \left< \tr_{U(n_c)}\frac{W_\alpha}{z-\Phi} \right>, \label{loop2} \\
&\frac{1}{64\pi^2} \left< \tr_{U(n_c)}\frac{W^\alpha W_\alpha}{z-\Phi} \right>=
\left(\frac{1}{64\pi^2} \left< \tr_{U(n_c)}\frac{W^\alpha W_\alpha}{z-\Phi} \right> \right)^2. \\
&n_f \frac{1}{64\pi^2} \left< \tr_{U(n_c)}\left( \frac{W^\alpha W_\alpha}{z-\Phi}\right) \right>
= \left<\tr_{U(n_f)}\left( \tilde{Q}\frac{m(\Phi)}{z-\Phi}Q \right) \right>.
\end{align}
These equations correspond to that derived from the field theory in ref.\cite{S}.

\subsection{Properties of the prepotential}
\label{3.2}
In ref.\cite{S}, owing to the chiral ring properties, the fundamental matter contributes to the effective action only up to the one-loop order.
We will show that the supermatrix model reproduces this property.

We first integrate out $\hat{Q}$ and $\hat{\tilde{Q}}$ in the partition function,
\begin{align}
\exp \left( -\frac{\hat{N}^2}{g_m^2}F_m \right)
=&  \int d \hat{\Phi}  \left( \det\frac{\hat{N}}{g_m} m(\hat{\Phi}) \right)^{-\Str_{U(\hat{n}_f)}(1)}  \exp\left(- \frac{\hat{N}}{g_m}\Str_{U(\hat{n}_c)}W(\hat{\Phi}) \right) \nonumber \\
=&  \int d \hat{\Phi} e^{- S_{eff}}, \nonumber \\
S_{eff}& = \frac{\hat{N}}{g_m}\left( \Str_{U(\hat{n}_c)}W(\hat{\Phi}) + \frac{g_m}{\hat{N}}\Bigl(\Str_{U(\hat{n}_f)}1\Bigr) \log \left(\det\frac{\hat{N}}{g_m} m(\hat{\Phi})\right)\right).   
\end{align}
We define here $S_{eff}$ as an effective action of $\hat{\Phi}$.
$\Str_{U(\hat{n}_f)}1$ is a one-loop correction of the fundamental matter. 
We can expand $F_m$ with respect to $\frac{g_m}{\hat{N}} \left< \Str_{U(\hat{n}_f)} 1 \right> $ as
\begin{align} 
F_m = F_{m0} + \sum_{k=1} \left( \frac{g_m}{\hat{N}}\left< \Str_{U(\hat{n}_f)}1 \right> \right)^k F_{mk}.
\label{pre2}
\end{align}
Note that the $F_{mk}$ is the $k$-loop  correction for the fundamental matter. 
We can now map this expansion to the gauge theory, and then map the free energy to the prepotential.
In addition, 
\begin{align}
 \frac{g_m}{\hat{N}} \left< \Str_{U(\hat{n}_f)}1 \right> =& \frac{\hat{N}}{g_m} \left< \Str_{U(\hat{n}_f)}\left( \delta^4(\hat{x}-x)\delta^2(\hat{\theta}-\theta) \right)^2 \right> \nonumber \\
 \mapsto& \frac{1}{64\pi^2} \left< \tr_{U(n_f)} ( W^\alpha_f -8\pi\psi^\alpha )(W_{f\alpha}-8\pi\psi_\alpha) \right>,
 \label{Konishi2}
 \end{align}
 where we take the limit, $C, \gamma \rightarrow 0$, and use the regularization of Eq.(\ref{Konishi}) for the $U(n_f)$ gauge field.
 By using these relationships, when we take $W_{f\alpha}=0$, Eq.(\ref{pre2}) is mapped to
\begin{align}
\mathcal{F} = \mathcal{F}_0 + n_f (\psi^\alpha \psi_\alpha) \mathcal{F}_1.
\end{align}
We will map $F_{mk}$ to  $\mathcal{F}_k$.
The anti-commutativity of $\psi_\alpha$ removes  $\mathcal{F}_k$  $(k>1)$.
The first term is the contribution of the adjoint matter, and the second is from the adjoint matter and one-loop correction of the fundamental matter.
The prepotential obtained from the free energy of the supermatrix model thus also has the contribution of the fundamental matters only up to the one-loop order.
This statement is consistent with the gauge theory analysis.\\

With the argument of the previous subsection, our supermatrix model gives the same result in the holomorphic parts of the field theory.  

We apply the theory with single fundamental matter in this section.
We can generally construct a corresponding supermatrix model in the gauge theory with several unitary gauge groups and fundamental and bifundamental matter as described in sections 2 and 3.

\section{Discussion}
\label{4}
\setcounter{equation}{0}
We have shown that the Dijkgraaf-Vafa theory in the $\mathcal{N}=1$ gauge theory with fundamental and bifundamental matter can be regarded as large-$N$ reduction.
This enables us to understand the mechanism by which the matrix model incorporates the information of the gauge theory. 

We have also shown the mapping from the gauge theories to corresponding supersymmetric reduced models.
However, there are some problems in the analyses of low-energy gauge theory.
For example, if gauge theory is coupled to massless fundamental matter, 
the low-energy theory should be described by meson and baryon fields \cite{ADS}.
We need to introduce these fields through the Legendre transformation,
in which case,  the mapping itself cannot prescribe them.  

Our approach cannot yet be applied to other gauge groups:- $SO(N)$ or $Sp(N)$.
It is difficult to construct the reduced model for such gauge theories,
in which the proof of equivalence between the gauge theories and matrix model is given \cite{W}.
Success with the Dijkgraaf-Vafa theory  might shed some light on constructing a reduced model in these gauge theories.

\begin{center} \begin{large}
Acknowledgments
\end{large} \end{center}
We thank T. Azuma, N. Ishibashi, S. Kawamoto, T.Kobayashi and T. Yokono 
for useful discussions. 
We are grateful to D. Klemm for informing us of their paper \cite{NCS} in which they also point out the non-commutative superspace.
We are grateful to R. Argurio for informing us of their paper \cite{Hof} in which they also point out the application of Dijkgraaf-Vafa techniques to theories with fundamental flavors.
We are grateful to N. Sadooghi for informing us of their paper \cite{KS} in which they consider various anomalies in non-commutative space.
The work of T.K. was supported in part by a JSPS research fellowships for young scientists.

\appendix
\section{A Construction of the reduced model}
\setcounter{equation}{0}
We summarize in this appendix the method for constructing the supermatrix model from the gauge theory.
\subsection{From the commutative theory to non-commutative theory}
\label{a1}
\paragraph{Non-commutative space\\}
We define the $*$- and $\star$-products as
\begin{align}
f(x)*g(x)=\left. \exp \left(-\frac{i}{2}C^{\mu\nu}\frac{\partial}{\partial x^\mu} \frac{\partial}{\partial y^\nu} \right)f(x)g(y) \right|_{y=x},\label{star1} \\
f(\theta)\star g(\theta)=\left. \exp \left(-\frac{1}{2}\gamma^{\alpha\beta}\frac{\partial}{\partial \theta^\alpha} \frac{\partial}{\partial {\theta'}^\beta} \right)f(\theta)g(\theta') \right|_{\theta=\theta'},
\label{star2} 
\end{align}
where $C^{\mu\nu}$ is an anti-symmetric tensor and $\gamma^{\alpha\beta}$ is a symmetric tensor.
The non-commutative coordinates then satisfy
\begin{align}
[ x^\mu , x^\nu]_* &= -i C^{\mu\nu}, \label{bosc} \\
\{ \theta^\alpha, \theta^\beta \}_\star &= \gamma^{\alpha\beta}.
\label{ferc}
\end{align}
\paragraph{Correspondence between  functions in non-commutative space and matrices \\}
The matrices corresponding to  bosonic non-commutative coordinates (Eq.(\ref{bosc})) are
\begin{align}
[ \hat{x}^\mu , \hat{x}^\nu] = -i C^{\mu\nu}.
\label{bosm}
\end{align}
In order to satisfy these relationships, the rank of these matrices, $\hat{N}$, should be  infinite.
We can introduce an anti-symmetric tensor which satisfies
\begin{align}
C^{\mu\lambda}B_{\lambda\nu}={\delta^\mu}_\nu
\end{align}
and define
\begin{align}
\hat{p}_\mu = B_{\mu\nu}\hat{x}^\nu.
\label{bosm2}
\end{align}
These matrices satisfy the following relations:
\begin{align}
[ \hat{p}_\mu , \hat{p}_\nu] = iB_{\mu\nu},~~[\hat{x}^\mu,\hat{p}_\nu]=i \delta^\mu_\nu.
\end{align}
Matrices corresponding to the fermionic coordinates (Eq.(\ref{ferc})) are
\begin{align}
\{ \hat{\theta}^\alpha, \hat{\theta}^\beta \} = \gamma^{\alpha\beta}.
\label{fer}
\end{align}
We can then introduce $\beta_{\alpha\beta}$ and $\pi_{\alpha}$ which satisfy
\begin{align}
\{\hat{\pi}_\alpha,\hat{\pi}_\beta \}=\beta_{\alpha\beta},~~ \{ \hat{\theta}^\alpha,\hat{\pi}_\beta \}=\delta^\alpha_\beta.
\end{align}
The correspondence between the functions and the matrices from these relationships is
\begin{align}
O(x)&=\int \frac{d^4 k}{(2\pi)^4} e^{ik_\mu x^\mu} \tilde{O}(k) &\mapsto~~~~ \hat{O}
&=\int \frac{d^4 k}{(2\pi)^4} e^{ik_\mu \hat{x}^\mu} \tilde{O}(k), 
\label{Weyl2} \\
Q(\theta)&=\int d^2 \kappa~e^{i\theta^\alpha \kappa_\alpha}\tilde{Q}(\kappa) 
&\mapsto~~~~ \hat{Q}
&=\int d^2\kappa~  e^{i\hat{\theta}^\alpha \kappa_\alpha}\tilde{Q}(\kappa) \nonumber \\
&= A + \theta^\alpha \psi_\alpha -(\theta^1\star\theta^2-\theta^2\star\theta^1) F  &
&= A + \hat{\theta}^\alpha \psi_\alpha - (\hat{\theta}^1\hat{\theta}^2-\hat{\theta}^2\hat{\theta}^1)F
\label{Weyl}
\end{align}
\paragraph{Correspondence of operations in bosonic coordinates \\}
The integral in the bosonic non-commutative coordinates is equivalent to the trace of the matrix model,
\begin{align}
\int d^4x~ O(x) 
&=(2\pi)^2\sqrt{\det C}\Tr_{U(\hat{N})}( \hat{O} ).
\end{align}
If $\hat{p}_\mu$ is a reducible representation and can represent $\hat{p}_\mu=\hat{p}_\mu^{(0)}\otimes 1_n $, where $\hat{p}_\mu^{(0)}$ is a certain irreducible representation of Eq.(\ref{bosm2}), then the relationship becomes
\begin{align}
\int d^4x~\tr_{U(n)}O(x)=(2\pi)^2\sqrt{\det C}\Tr_{U(\hat{N})}( \hat{O} ).
\label{map10}
\end{align}
On the other hand, the derivative corresponds as follows: 
\begin{align}
-i\partial_\mu O(x) &\mapsto [\hat{p}_\mu,\hat{O}]. 
\label{map3} 
\end{align}
\paragraph{Correspondence of operations in fermionic coordinates \\}
By using the $SL(2,C)$ transformation, $\gamma^{\alpha\beta}$ can be taken as
\begin{align}
(\gamma^{\alpha\beta})
=
\begin{pmatrix}
\gamma & 0 \\
0 & \gamma 
\end{pmatrix}.
\end{align}
In this case, $\hat{\theta}^\alpha$ can be represented in terms of Pauli matrices as
\begin{align}
\hat{\theta}^1 = \sqrt{\gamma}\sigma^1,~~\hat{\theta}^2=\sqrt{\gamma}\sigma^2, \nonumber \\
\hat{\theta}^1\hat{\theta}^2-\hat{\theta}^2\hat{\theta}^1 = 2i \gamma \sigma^3.
\end{align}
An integral for the fermionic non-commutative coordinates is
\begin{align}
F=\int d^2\theta Q(\theta) \mapsto  F &=\frac{i}{8\gamma} \tr \left(2 \sigma^3 \hat{Q} \right)  \nonumber \\
&\equiv \frac{i}{8\sqrt{\det \gamma }} \Str \left( \hat{Q} \right),
\label{Str}
\end{align}
where we define the supertrace, $\Str \hat{Q}= \tr 2\sigma^3\hat{Q}$.
Considering a quantum degree of the freedom of the non-commutative field theory, we can put
\begin{align}
\frac{i(2\pi)^2\sqrt{\det C}}{8 \sqrt{\det \gamma}}=\frac{\hat{N}}{g_m}.
\label{coefficient}
\end{align}
Here, $g_m$ is an appropriate constant with mass dimension 3.
As a result, we can obtain
\begin{align}
\int d^4x d^2\theta O(x,\theta) = \frac{\hat{N}}{g_m} \Str_{x\otimes\theta}\hat{O}.
\label{integral}
\end{align}\\

A derivative is mapped  as follows:
\begin{align}
\frac{\partial}{\partial \theta^\alpha}O(x,\theta) &\mapsto [ \hat{\pi}_\alpha,\hat{O} \}. \label{map4}
\end{align}
\subsection{Construction of the supersymmetric reduced model}
\label{a2}
We show the construction of the supersymmetric reduced model of an $\mathcal{N}=1$ supersymmetric $U(n_c)$ gauge theory coupled to adjoint matter $\Phi$.
We start from the action,
\begin{align}
S=&\int d^4xd^2\theta d^2\bar{\theta}~ \tr_{U(n_c)}(\bar{\Phi}e^V \Phi e^{-V}) \nonumber \\
+ &\int d^4x d^2\theta~2\pi i\tau~\tr_{U(n_c)}(W^\alpha W_\alpha)
+ \int d^4x d^2\theta~ \tr_{U(n_c)}(W(\Phi)) + c.c..
\label{action6}
\end{align}
We can then rewrite the fields in an appropriate form for the reduced model as follows:
\begin{align}
\Phi(x,\theta,\bar{\theta})=&\Phi^{(y)} (y,\theta) \nonumber \\
=&e^A \Phi^{(y)}(x,\theta) e^{-A}, \label{shift1} \\
\bar{\Phi}(x,\theta,\bar{\theta})=&\bar{\Phi}^{(y^\dagger)}(y^\dagger,\bar{\theta})\nonumber \\
=& e^{-A} \bar{\Phi}^{(y^\dagger)}(x,\bar{\theta}) e^A, \label{shift2} \\
\mathcal{D}_\alpha =& e^{A} D_\alpha e^{-A} = \frac{\partial}{\partial \theta^\alpha}, \label{shift3} \\
\bar{\mathcal{D}}_{\dot{\alpha}} =& e^{-A} \bar{D}_{\dot{\alpha}}e^{A} = - \frac{\partial}{\partial \bar{\theta}^{\dot{\alpha}}}, \label{shift4} \\
e^{\cV(x,\theta,\bar{\theta})}=&e^A e^{V(x,\theta,\bar{\theta})} e^A, 
\label{shift5} \\
W^{(y)}_\alpha(x,\theta) =& e^{-A} W_\alpha(x,\theta)e^A \nonumber \\
=& -\frac{1}{4}\bar{\mathcal{D}} \bar{\mathcal{D}} e^{-\cV} \mathcal{D}_\alpha e^{\cV},
\label{shift6}
\end{align}
where $A$ is the differential operator, $i\theta \sigma^\mu \bar{\theta} \partial_\mu$, and subscript $(y)$ ( or  $(y^\dagger)$ ) means that the chiral (or anti-chiral) superfield is a function of $(y=x+i\theta \sigma \theta, \theta)$ ( or $(y^\dagger,\bar{\theta})$ ).

We can then obtain the action,
\begin{align}
S=&\int d^4xd^2\theta d^2\bar{\theta}~ \tr_{U(n_c)}(\bar{\Phi}^{(y^\dagger)}e^{\cV} \Phi^{(y)} e^{-\cV}) \nonumber \\
+ &\int d^4x d^2\theta~2\pi i\tau~\tr_{U(n_c)}(W^{(y)\alpha} W_\alpha^{(y)})
+ \int d^4x d^2\theta~ \tr_{U(n_c)}(W(\Phi^{(y)})) + c.c..
\end{align}
We next consider this action on non-commutative space in Eqs.(\ref{bosc}) and (\ref{ferc}), and apply the mapping of Sec.\ref{a1}.
The action then becomes
\begin{align}
S_{rm}=& \frac{i^2(2\pi)^2\sqrt{\det C}}{8^2 \sqrt{\det \gamma}\sqrt{\det \gamma^*}} \Str_{U(\hat{n}_c)}\left( \hat{\bar{\Phi}}e^{\hat{\cV}}\hat{\Phi}e^{-\hat{\cV}}\right) \nonumber \\
&+ \frac{i(2\pi)^2\sqrt{\det C}}{8 \sqrt{\det \gamma}}\{ 2\pi i \tau \Str_{U(\hat{n}_c)}(\hat{W}^\alpha \hat{W}_\alpha) + \Str_{U(\hat{n}_c)}(W(\hat{\Phi}))\}+ c.c.,
\label{action5}
\end{align}
where $U(\hat{n}_c)$ indicates that $\hat{p}_\mu$ is the reducible representation, $\hat{p}_\mu=\hat{p}_\mu^{(0)} \otimes 1_{n_c}$.
This is the supersymmetric reduced model.

As a result of the holomorphy, we consider only the relevant parts in the Dijkgraaf-Vafa theory,
\begin{align} 
S_{smm}=\frac{\hat{N}}{g_m} \Str_{U(\hat{n}_c)} W(\hat{\Phi}).
\label{smm}
\end{align}
This is an action of the supermatrix model corresponding to Eq.(\ref{action6}).

\end{document}